\documentclass{IEEEojcsys}

\usepackage[colorlinks,urlcolor=blue,linkcolor=blue,citecolor=blue]{hyperref}

\usepackage{color,array}

\usepackage{graphicx}

\usepackage{amsmath,amsfonts}
\usepackage{algorithmic}
\usepackage{algorithm}
\usepackage{array}
\usepackage[caption=false,font=normalsize,labelfont=sf,textfont=sf]{subfig}
\usepackage{textcomp}
\usepackage{stfloats}
\usepackage{url}
\usepackage{verbatim}
\usepackage{graphicx}
\usepackage{cite}
\usepackage{lipsum}
\usepackage{hyperref}
\usepackage{tikz}
\usepackage{amssymb}
\usepackage{tablefootnote}
\usepackage{longtable}
\usepackage{listings}
\usepackage{scalerel}
\usepackage{graphicx}
\usepackage{cases}
\usepackage{lineno}

\jvol{00}
\jnum{XX}
\paper{1234567}
\pubyear{202x}
\receiveddate{XX Y 202x}
\accepteddate{XX Y 202x}
\publisheddate{XX Y 202x}
\currentdate{XX Y 202x}
\doiinfo{OJCSYS.202x.Doi Number}

\begin{document}

\sptitle{Article Category}

\title{Control Barrier Function-Based Quadratic Programming for Safe Operation of Tethered UAVs} 


\author{S.O FOLORUNSHO\affilmark{1} (Student Member, IEEE)}

\author{M. NI\affilmark{2}}

\author{W.R NORRIS\affilmark{3}  (Member, IEEE)}


\affil{Department of Industrial and Enterprise Systems Engineering, University of Illinois, Urbana-Champaign, 61801 USA} 
\affil{Department of Aerospace Engineering, University of Illinois, Urbana-Champaign, 61801 USA} 
\affil{Department of Industrial and Enterprise Systems Engineering, University of Illinois, Urbana-Champaign, 61801 USA} 

\corresp{CORRESPONDING AUTHOR: S. O. Folorunsho (e-mail: \href{mailto:sof3@illinois.edu}{sof3@illinois.edu})}
\authornote{This work received no external funding}

\markboth{PREPARATION OF PAPERS FOR IEEE OPEN JOURNAL OF CONTROL SYSTEMS}{S. O. FOLORUNSHO {\itshape ET AL}.}

\begin{abstract}
Consider an unmanned aerial vehicle (UAV) physically connected to the ground station with a tether operating in a space, tasked with performing precise maneuvers while constrained by the physical limitation of its tether, which prevents it from flying beyond a maximum allowable length. Violating this tether constraint could lead to system failure or operational hazards, making it essential to enforce safety constraints dynamically while ensuring the drone can track desired trajectories accurately. This paper presents a Control Barrier Function Quadratic Programming Framework (CBF-QP) for ensuring the safe and efficient operation of tethered unmanned aerial vehicles (TUAVs). The framework leverages nominal backstepping control to achieve trajectory tracking, augmented with control barrier functions to ensure compliance with the tether constraint. In this proposed method, the tether constraint is directly embedded in the control design and therefore guarantees the TUAV remains within a predefined operational region defined by the maximum tether length while achieving precise trajectory tracking.

The effectiveness of the proposed framework is validated through simulations involving set-point tracking, dynamic trajectory following, and disturbances such as incorrect user inputs. The results demonstrate that the TUAV respects the tether constraint \(||\mathbf{x}(t)|| \leq L_{\text{max}}\), with tracking errors converging to zero and the control input remaining bounded.
\end{abstract}


\maketitle

\section{INTRODUCTION}

{T}ethered Unmanned Aerial Vehicles (TUAVs) have emerged as a promising solution for applications that require extended flight duration, increased stability, and robust connectivity, especially in areas where traditional battery-powered drones may fail. Applications of tethered drones span a wide range of fields, including surveillance, environmental monitoring, communication relays, and emergency response, where prolonged and reliable aerial presence is critical \cite{Marques_2023}.

With increasing emphasis on autonomous systems, safety has been brought to the forefront of design considerations \cite{Lamport_1977}, calling for rigorous safe control methods \cite{Hsu_2024, Wabersich_2023}. Ensuring safe and stable flight performance in tethered drone systems, particularly under varying conditions \cite{Koopman_2019}, remains a significant challenge. Safety principles are essential in the design of tethered drones, since these UAVs often operate near human operators \cite{Shi_2024}, other equipment, or infrastructure. Safety considerations include ensuring the drone does not invade restricted areas, maintaining a safe distance from obstacles \cite{Ding_2024}, and controlling the tension of the tie to prevent sudden or hazardous movements. Additionally, constraints on the drone’s position, velocity, and altitude due to the tether require advanced control strategies to ensure safe operation within predefined limits.

Control Barrier Functions (CBFs) \cite{Ames_2019} offer a novel approach to addressing these challenges. CBFs are mathematical tools designed to enforce safety constraints by regulating the behavior of the system, ensuring that critical boundaries are not crossed. In tethered drones, CBFs can be applied to maintain safety margins regarding altitude, proximity to obstacles, and tension limits on the tether. This allows the drone to respond dynamically to environmental disturbances or control inputs without compromising safety or stability. The recent development of control barrier functions indicates that many control design techniques, originally based on Lyapunov and control Lyapunov functions \cite{Li_2023}, can be adapted to prioritize safety requirements; thus providing a set of tools that integrate safety into control system design with a level of rigor comparable to that of liveness~\cite{Ames_2019}.

To effectively implement CBFs, Control Barrier Function Quadratic Programs (CBF-QPs) are employed \cite{Ames_2019, Ames_2014, Li_2023}. CBF-QPs enable the real-time enforcement of multiple safety constraints by solving a quadratic optimization problem that balances safety and control objectives \cite{Ames_2017}. In this framework, a CBF-QP formulates a quadratic program that takes into account the current state of the drone, the desired control inputs, and the predefined safety constraints. The objective is to minimize the deviation from the nominal control input while satisfying the CBF constraints, thus ensuring safe operation without sacrificing control performance. Through this QP constraint adjustment, operators can prioritize specific safety requirements, such as maintaining a minimum altitude or staying within a restricted flight zone. This flexibility enables a robust control architecture that can adapt to changing conditions and maintain safety in real time \cite{Rauscher_2016}.

\section{RELATED WORKS}

While CBFs have demonstrated their robustness and scalability \cite{Li_2021, Lopez_2021, Lindemann_2024} as a safety method, there is limited literature on their application in Unmanned Aerial Vehicles (UAVs) and to the best of available knowledge, almost none specifically for TUAVs. The application of CBFs in non-tethered UAVs offers valuable insights for managing stability, safety, and constraint adherence in tethered drone systems. In \cite{Singletary_2022}, an implicit CBF for high-speed safety provides a foundation for adapting spatial and safety constraints to confined environments. Studies by \cite{Panja_2023} and \cite{Tayal_2024} showcase collision avoidance techniques that could aid in managing tether proximity risks. \cite{Zheng_2023} shows an adaptive CBF for path-following under disturbances and \cite{Wang_2018, Wang_2023, Liu_2024} illustrates adaptive safe stabilization and event-triggered control approaches that further address environmental constraints and efficient control adaptations, vital for tethered UAVs. 

Other safety filters such as Model Predictive Control (MPC) \cite{Korsarnovsky_2020, Singh_2001, Lindqvist_2020}, Explicit Reference Governor (ERG) \cite{Hermand_2018}, reachability analysis \cite{Ankit_2024, Zhou_2015, Ding_2012} have been applied to UAVs. However, they lack the real-time adaptability and formal safety guarantees provided by Control Barrier Functions (CBFs). CBFs have been shown to be more computationally efficient \cite{Li_2021} and less conservative in enforcing safety constraints \cite{Tayal_2024}. Most recent research on TUAV safety filters is largely based on Model Predictive Control (MPC) \cite{Bolognini_2022, Valerio_2022}; however, MPC can be computationally expensive and may not always guarantee real-time performance, particularly for systems with fast dynamics or stringent computational constraints such as the TUAV. CBFs offer a flexible and adaptable framework that allows for dynamic adjustment of safety boundaries as conditions change, which is crucial in TUAV systems that must account for both tether length and external environmental factors. 

Motivated by the societal need for safety in TUAV operations, this paper explores the integration of Control Barrier Function Quadratic Programs into the control architecture of tethered drones to enhance robot safety and operational efficiency. The specific contributions of this paper are as follows: the development of a nonlinear backstepping control approach for the precise control of tethered unmanned aerial vehicles (TUAVs) in three-dimensional space; the explicit integration of backstepping control with Control Barrier Functions (CBFs) to enforce safety constraints dynamically; the utilization of Quadratic Programming (QP) to ensure constraint satisfaction and optimize control inputs in real time; and the establishment of a flexible control framework that can be extended to other tethered systems or safety-critical robotic applications. 

\section{System Description}
\label{sysdes}

\subsection{Tethered Unmanned Aerial Vehicle}
Consider a system of a tethered Unmanned Aerial Vehicle (UAV) described by the following equations of motion \ref{dynamics} as in \cite{folorunsho2024nonlinearcontrolstabilityanalysis, liu2021dynamic} and represented geometrically as in Fig.\ref{system_des} constrained within an hemispherical flight region:

\begin{figure}[ht]
\centering
\includegraphics[width=0.5\textwidth]{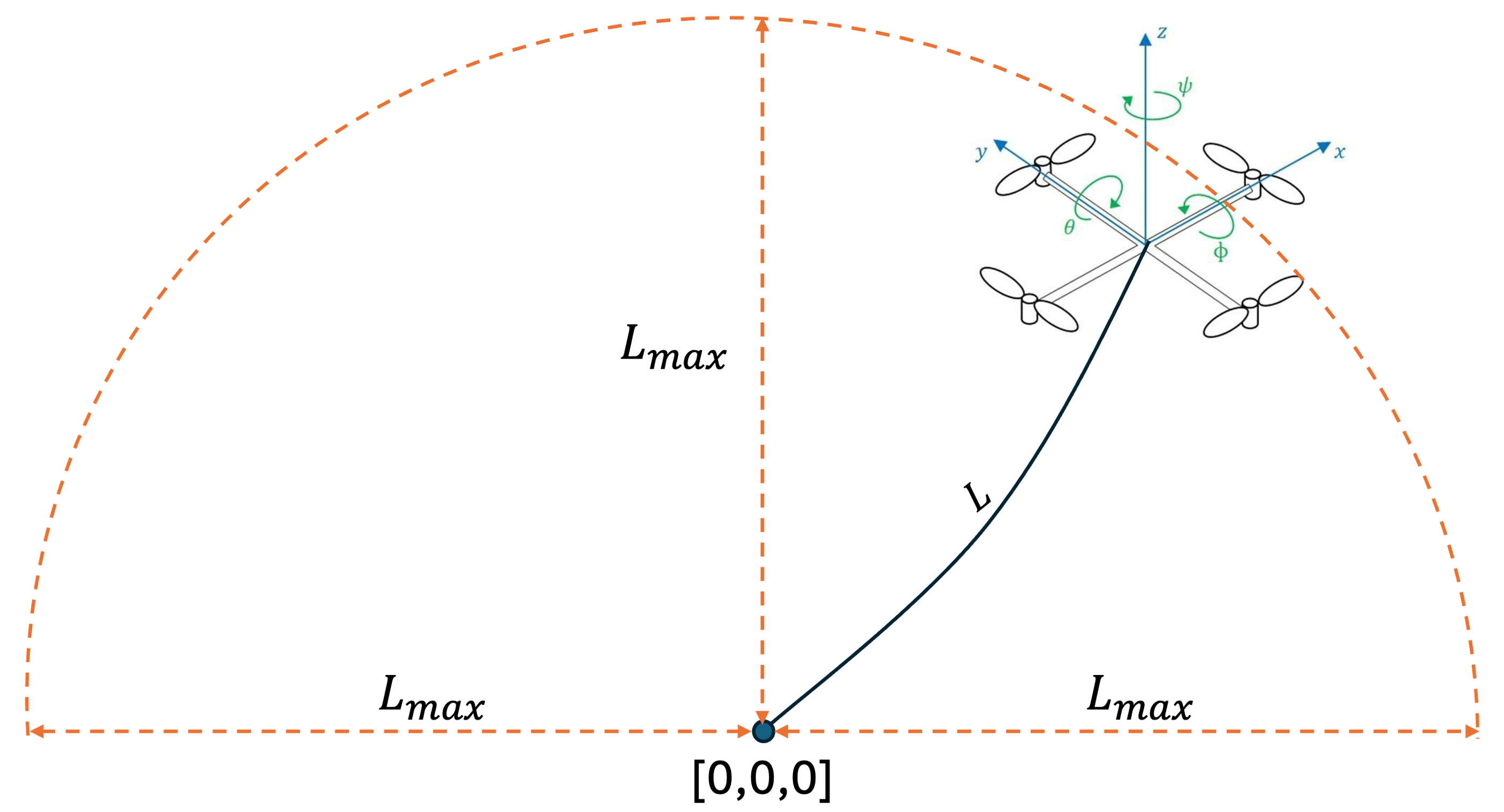}
\caption{Illustration of the flight region of a tethered UAV, constrained within a hemispherical volume defined by the tether length. The inset shows the UAV's orientation dynamics.}
\label{system_des}
\end{figure}

\renewcommand{\arraystretch}{1.5} 
\begin{equation}
\label{dynamics}
    \begin{cases}
        \ddot{x} = \frac{U_f (C_\psi C_\phi S_\theta + S_\psi S_\phi) + mrv - mqw}{m} \\
        \quad - \frac{mgS_\theta + T_1 C_\alpha S_\beta + A_{x}u}{m}, \\[10pt]
        \ddot{y} = \frac{U_f (C_\phi S_\psi S_\theta - C_\psi S_\phi) + mru - mpw}{m} \\
        \quad - \frac{mgC_\theta S_\phi - T_1 C_\alpha C_\beta - A_{y}v}{m}, \\[10pt]
        \ddot{z} = \frac{U_f C_\theta C_\phi + mqu - mpv + mgC_\theta C_\phi - T_1 S_\alpha - A_{z}w}{m}, \\[10pt]
        \ddot{\phi} = \frac{U_\phi - qr(I_{yy} - I_{zz})}{I_{xx}}, \\[10pt]
        \ddot{\theta} = \frac{U_\theta + pr(I_{xx} - I_{zz})}{I_{yy}}, \\[10pt]
        \ddot{\psi} = \frac{U_\psi - pq(I_{xx} - I_{yy})}{I_{zz}}, \\[10pt]
        \ddot{\vartheta} = \frac{1}{I_w} \left( - \beta_w \dot{\vartheta} + r_w U_{win} \right),
    \end{cases}
\end{equation}

where \( (x, y, z) \in \mathbb{R}^3 \) are the Cartesian positions of the UAV center of mass in the inertial frame, and \( (\phi, \theta, \psi) \in \mathbb{R}^3 \) represent the roll, pitch, and yaw angles. The state velocities \( (u, v, w) = (\dot{x}, \dot{y}, \dot{z}) \) and \( (p, q, r) = (\dot{\phi}, \dot{\theta}, \dot{\psi}) \) correspond to the linear and angular velocities, respectively.

The system includes contributions from:
\begin{itemize}
    \item \textbf{Control inputs:} \( U_f, U_\phi, U_\theta, U_\psi \), which represent thrust, roll, pitch, and yaw inputs.
    \item \textbf{Tether tension:} \( T_1 \), resolved into components along the inertial axes through angles \( \alpha \) and \( \beta \), defined as in \cite{liu2021dynamic}:
    \begin{align*}
        T_X &= T_1 \cos \alpha \sin \beta, \\
        T_Y &= T_1 \cos \alpha \cos \beta, \\
        T_Z &= T_1 \sin \alpha.
    \end{align*}
    \item \textbf{Drag forces:} \( A_x u, A_y v, A_z w \) act along the \( x, y, z \)-axes.
    \item \textbf{Winch dynamics:} The tether is controlled via the winch torque \( U_{win} \), with dynamics governed by $\ddot{\vartheta}$ in equation \ref{dynamics} where \( \vartheta \) represents the angular position of the winch.
\end{itemize}

The UAV's rotational dynamics depend on its moments of inertia \( I_{xx}, I_{yy}, I_{zz} \) and angular rates, with coupling introduced through gyroscopic effects.

\subsection{Control Barrier Function}

Control Barrier Functions (CBFs) provide a systematic framework for ensuring the safety of dynamical systems by enforcing state constraints. The concept revolves around the forward invariance of a set, ensuring that the system's state trajectory remains within a predefined safe set over time. This property is critical for tethered UAV systems, where the position must always respect constraints imposed by the tether.

Consider the control-affine system dynamics:
\begin{equation}
    \dot{x} = f(x) + g(x)u, \quad x \in \mathbb{R}^n, \; u \in \mathbb{R}^m,
\end{equation}
where \( f(x) \) and \( g(x) \) are locally Lipschitz continuous, \( x \) represents the system state, and \( u \) is the control input.

Let the safe set \( \mathcal{C} \subset \mathbb{R}^n \) be defined as:
\begin{equation}
    \mathcal{C} = \{x \in \mathbb{R}^n \mid h(x) \geq 0 \},
\end{equation}
where \( h(x): \mathbb{R}^n \to \mathbb{R} \) is a continuously differentiable function. The boundary and interior of \( \mathcal{C} \) are given by:
\[
    \partial\mathcal{C} = \{x \in \mathbb{R}^n \mid h(x) = 0\},
\]
\[
    \text{Int}(\mathcal{C}) = \{x \in \mathbb{R}^n \mid h(x) > 0\}.
\]

The set \( \mathcal{C} \) is forward invariant if, for any \( x(0) \in \mathcal{C} \), the solution \( x(t) \in \mathcal{C} \) for all \( t \geq 0 \). Forward invariance ensures that the system remains within the safe set indefinitely.

A function \( h(x) \) is called a Control Barrier Function (CBF) if there exists an extended class \( \mathcal{K}_\infty \) function \( \alpha \) such that for all \( x \in \mathcal{C} \):
\begin{equation}
    \sup_{u \in \mathbb{R}^m} \left[ L_f h(x) + L_g h(x) u \right] \geq -\alpha(h(x)),
\end{equation}
where \( L_f h(x) = \frac{\partial h(x)}{\partial x} f(x) \) and \( L_g h(x) = \frac{\partial h(x)}{\partial x} g(x) \) are the Lie derivatives of \( h(x) \) with respect to \( f(x) \) and \( g(x) \), respectively.

The inequality ensures that the set \( \mathcal{C} \) is forward invariant under the control law \( u \). In this work, the CBF is designed to ensure that the UAV position remains within the constraints imposed by the tether length and physical limitations of the system.

\subsection{Control Objectives}
The primary control objective is to design a safe and robust control strategy for the tethered UAV system, ensuring the UAV's position remains within the constraints imposed by the tether while maintaining stability and achieving desired performance. Using backstepping control as the base control method, a Quadratic Program (QP) is formulated to integrate Control Barrier Functions (CBFs) for enforcing safety-critical constraints. This approach ensures forward invariance of a predefined safe set while respecting the dynamic interactions between the UAV, tether, and winch.

\section{Controller design and stability analysis}
\label{control}
\begin{figure}[ht]
\centering
\includegraphics[width=0.5\textwidth]{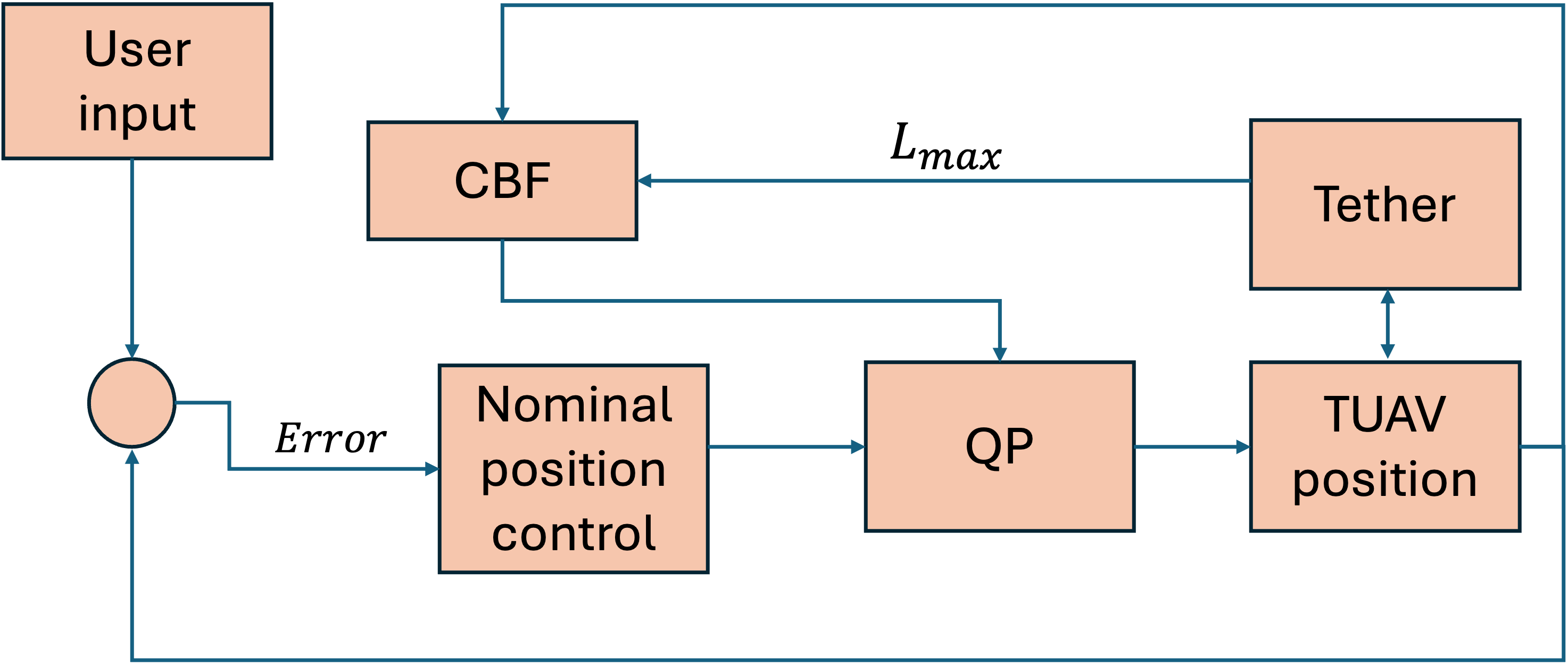}
\caption{Overall Control Architecture.}
\label{control_arch}
\end{figure}

In this section, the backstepping control law is developed as the baseline controller for the tethered UAV system, ensuring stability and performance. Building on this, a Control Barrier Function (CBF) is designed to enforce safety constraints, and a Quadratic Program (QP) is formulated to integrate the CBF with the control law. This approach guarantees that the UAV operates within the constraints imposed by the tether while achieving safe and robust performance. The overall control architecture is shown in Fig.~\ref{control_arch}.

\subsection{Backstepping Control}

The backstepping control methodology is employed to ensure the stability and trajectory tracking of the UAV's dynamics. For illustrative purposes, this section focuses on the altitude control in the \( z \)-direction.

The altitude dynamics are expressed as:
\begin{equation}\label{mainsys1}
\begin{aligned}
\dot{x}_5 &= x_6 \\
\dot{x}_6 &= \frac{1}{m}\left( U_f \cos x_9 \cos x_7 +
m x_{10} x_{12} - m x_8 x_4 \right) \\
&\quad + \frac{1}{m}\left( mg \cos x_9 \cos x_7 + T_z - A_z x_6 \right)
\end{aligned}
\end{equation}
where \( x_5 \) represents the UAV's \( z \)-position, and \( x_6 \) its vertical velocity. The control input \( U_f \) represents the thrust force, and other terms account for system dynamics such as drag (\( A_z x_6 \)) and tether tension (\( T_z \)).

The altitude controller is designed using the following approach:

\paragraph{Step 1: Virtual control design.}
A Lyapunov function is proposed to analyze the system's stability:
\[
V_1 = \frac{1}{2}x_5^2,
\]
where \( V_1 > 0 \) for all \( x_5 \neq 0 \), and \( V_1 = 0 \) when \( x_5 = 0 \). Its time derivative is:
\[
\dot{V}_1 = x_5 \dot{x}_5 = x_5 x_6.
\]

To stabilize the system, a virtual control law is selected:
\[
\phi(x_5) = -k_1 x_5,
\]
where \( k_1 > 0 \) is a design parameter. Substituting \( \phi(x_5) \) ensures that \( \dot{V}_1 = -k_1 x_5^2 \), which is negative definite, proving that the system is asymptotically stable.

\paragraph{Step 2: State transformation.}
To extend the control design, a state transformation is introduced:
\[
z_1 = x_6 - \phi(x_5),
\]
where \( z_1 \) represents the deviation of the actual velocity \( x_6 \) from the virtual control. Taking its time derivative:
\[
\dot{z}_1 = \dot{x}_6 - \dot{\phi}(x_5),
\]
and substituting for \( \dot{x}_6 \) and \( \dot{\phi}(x_5) = -k_1 \dot{x}_5 = -k_1 x_6 \), the expression becomes:
\[
\dot{z}_1 = \frac{1}{m} \left( U_f \cos x_9 \cos x_7 + mx_{10}x_{12} - mx_8x_4 \right.
\]
\[
\left. + mg\cos x_9\cos x_7 + T_z - A_z x_6 \right) + k_1 x_6.
\]

\paragraph{Step 3: Composite Lyapunov function.}
To analyze the stability of the transformed system, a composite Lyapunov function is defined as:
\[
V_{c1} = V_1 + \frac{1}{2}z_1^2 = \frac{1}{2}x_5^2 + \frac{1}{2}z_1^2.
\]
The time derivative of \( V_{c1} \) is given by:
\[
\dot{V}_{c1} = \dot{V}_1 + z_1 \dot{z}_1 = x_5 z_1 - k_1 x_5^2 + z_1 \dot{z}_1.
\]

To ensure that \( \dot{V}_{c1} \) is negative definite, the auxiliary control law for \( \dot{z}_1 \) is chosen as:
\[
\dot{z}_1 = -k_2 z_1 - x_5,
\]
where \( k_2 > 0 \) is another design parameter. Substituting this choice into \( \dot{V}_{c1} \), resulted in:
\[
\dot{V}_{c1} = -k_1 x_5^2 - k_2 z_1^2,
\]
which is negative definite, ensuring the stability of the system.

\paragraph{Step 4: Final control input.}
Using the expression for \( \dot{z}_1 \), the thrust control input \( U_f \) is derived as:
\begin{align}
U_f &= \frac{-m x_{10} x_{12} + m x_8 x_4 - mg \cos x_9 \cos x_7 - T_z}{\cos x_9 \cos x_7} \notag \\
&\quad + \frac{A_z x_6 - m x_5 - m k_2 z_1 - m k_1}{\cos x_9 \cos x_7}.
\end{align}

This control input stabilizes the altitude dynamics and facilitates trajectory tracking. Similar procedures are followed for the \( x \)- and \( y \)-directions, attitude (\( \phi, \theta, \psi \)), and tether length control. The detailed derivations for these controllers are provided in the previous work of the authors \cite{folorunsho2024nonlinearcontrolstabilityanalysis}.

\subsection{Control Barrier Function for Tethered UAV System}

Following the development of the nominal control laws using the backstepping approach, a Control Barrier Function (CBF) is formulated to enforce safety-critical constraints on the tethered UAV system. The objective is to ensure that the UAV’s position remains within the physical limits imposed by the tether length while stabilizing the system dynamics.

Consider a tethered UAV system, whose states and inputs include the UAV’s position \(\boldsymbol{\xi} = (x,y,z)\), orientation angles \((\phi, \theta, \psi)\), winder angle \(\vartheta\), and their respective velocities as described in \ref{dynamics}. Let \(\chi \in \mathbb{R}^n\) represent the full state vector, and let
\[
u = [U_f, U_\phi, U_\theta, U_\psi, U_{\text{win}}]^T
\]
denote the control input vector, where \(U_f\) is the collective thrust, \(U_{\phi}, U_{\theta}, U_{\psi}\) are the attitude control inputs, and \(U_{\text{win}}\) is the winder torque input.

From the dynamics of the system, it is possible to express the system in a control-affine \cite{li2023survey} form:
\[
\dot{\chi} = f(\chi) + g(\chi)u.
\]
Here, \( f:\mathbb{R}^n \to \mathbb{R}^n \) and \( g:\mathbb{R}^n \to \mathbb{R}^{n \times 5} \) are locally Lipschitz functions, ensuring well-posedness of the closed-loop system trajectories. The functions \(f(\chi)\) and \(g(\chi)\) implicitly incorporate gravitational effects, aerodynamic terms, and any internal coupling in the UAV-winder system, while the input \(u\) directly influences the evolution of \(\chi\) through \(g(\chi)\).

Define the safety function based on the maximum allowable tether length \(L_{\text{max}}>0\):
\[
h(\chi) = L_{\text{max}} - \|\boldsymbol{\xi}\|, \quad \text{where } \|\boldsymbol{\xi}\| = \sqrt{x^2 + y^2 + z^2}.
\]
The safe set is:
\[
\mathcal{C} = \{\chi \in \mathbb{R}^n \mid h(\chi) \ge 0\}.
\]

A Control Barrier Function (CBF) requires an extended class-\(\mathcal{K}_\infty\) function \(\alpha\) such that for all \(\chi \in \mathcal{C}\):
\[
\sup_{u \in \mathbb{R}^5}[L_f h(\chi) + L_g h(\chi)u] \ge -\alpha(h(\chi)),
\]
where \(L_f h(\chi) = \nabla h(\chi)^T f(\chi)\) and \(L_g h(\chi) = \nabla h(\chi)^T g(\chi)\).

Since
\[
\nabla h(\chi) = -\frac{\boldsymbol{\xi}}{\|\boldsymbol{\xi}\|} \oplus 0,
\]
it follows that
\[
L_f h(\chi) = -\frac{\boldsymbol{\xi}^T f_{\text{pos}}(\chi)}{\|\boldsymbol{\xi}\|}, \quad L_g h(\chi) = -\frac{\boldsymbol{\xi}^T g_{\text{pos}}(\chi)}{\|\boldsymbol{\xi}\|},
\]
where \(f_{\text{pos}}(\chi)\) and \(g_{\text{pos}}(\chi)\) are the position-related components of \(f(\chi)\) and \(g(\chi)\). Substituting these into the CBF condition gives:
\[
L_f h(\chi) + L_g h(\chi)u \ge -\alpha(h(\chi))
\]
\[
-\frac{\boldsymbol{\xi}^T f_{\text{pos}}(\chi)}{\|\boldsymbol{\xi}\|} -\frac{\boldsymbol{\xi}^T g_{\text{pos}}(\chi)u}{\|\boldsymbol{\xi}\|} \ge -\alpha(h(\chi)),
\]
which can be rearranged as:
\[
\boldsymbol{\xi}^T g_{\text{pos}}(\chi) u \le \|\boldsymbol{\xi}\|\alpha(h(\chi)) + \boldsymbol{\xi}^T f_{\text{pos}}(\chi).
\]
\vspace{0.1in}
\noindent \textbf{Proof of Forward Invariance}: To show forward invariance of \(\mathcal{C}\), assume \(\chi(0)\in \mathcal{C}\) so that \(h(\chi(0))\ge0\). Along the trajectory:
\[
\dot{h}(\chi) = L_f h(\chi) + L_g h(\chi)u \ge -\alpha(h(\chi)).
\]

Consider the scalar differential inequality \(\dot{h}(t) \ge -\alpha(h(t))\), where \(h(t)=h(\chi(t))\). By the comparison lemma, if \(y(t)\) solves \(\dot{y}(t)=-\alpha(y(t))\) with \(y(0)=h(\chi(0))\ge0\), then \(h(t)\ge y(t)\) for all \(t\). Since \(\alpha\) is class-\(\mathcal{K}_\infty\), \(\alpha(\cdot)\ge0\) and \(\alpha(0)=0\), the solution \(y(t)\) cannot become negative. Hence, \(h(t)\) remains nonnegative for all time. This result also supports the geometric intuition provided by Nagumo’s theorem \cite{constantin2010nagumo}, as it ensures that the vector field at the boundary does not push the trajectories outside \(\mathcal{C}\).

Since \(h(\chi(t))\ge0\) for all \(t\ge0\), the trajectory remains in \(\mathcal{C}\), proving forward invariance.

To implement this in the TUAV system, the CBF constraint is integrated into a Quadratic Program (QP). Let \(u_{\text{nominal}}\) be a control input from a backstepping-based nominal controller described in \ref{control}. The QP is:
\[
\begin{aligned}
\min_{u} \quad & \tfrac{1}{2}\|u - u_{\text{nominal}}\|^2 \\
\text{subject to} \quad & \boldsymbol{\xi}^T g_{\text{pos}}(\chi) u \le \|\boldsymbol{\xi}\|\alpha(h(\chi)) + \boldsymbol{\xi}^T f_{\text{pos}}(\chi).
\end{aligned}
\]

This QP enforces that among all control inputs maintaining the safety constraint, the one closest to the nominal input is chosen, thereby ensuring both performance (through the nominal controller) and safety (through the CBF).

\section{Results and Simulations}
\label{results}

Consider the TUAV system described in Section~\ref{sysdes}. The simulation parameters are summarized in Table~\ref{tab:parameters}.

\begin{table}[htbp]
    \centering
    \caption{System and Tether Parameters}
    \label{tab:parameters}
    \begin{tabular}{ll}
        \hline
        \textbf{Parameter} & \textbf{Value} \\
        \hline
        Mass \(m\) & 2.84 kg \\
        Gravitational acceleration \(g\) & 9.81 m/s\(^2\) \\
        Moment of inertia \(I_{xx}\) & 0.5192 kg·m\(^2\) \\
        Moment of inertia \(I_{yy}\) & 0.4929 kg·m\(^2\) \\
        Moment of inertia \(I_{zz}\) & 0.0947 kg·m\(^2\) \\
        Tether density \(\rho\) & 0.034 kg/m \\
        Tether cross-sectional area \(A\) & \(1.1 \times 10^{-4}\) m\(^2\) \\
        Maximum tether length \(L_{\text{max}}\) & 13 m \\
        Winch effective radius \(r_w\) & 0.05 m \\
        Winch viscous friction \(\beta_w\) & 0.01 N·m·s/rad \\
        Winch moment of inertia \(I_w\) & 0.002 kg·m\(^2\) \\
        Tether stiffness constant \(K_t\) & 1000 N/m \\
        \hline
    \end{tabular}
\end{table}

The simulation results of the controllers developed using these parameters are presented in the following sections. The results demonstrate the effectiveness of the control strategy in maintaining system stability and ensuring compliance with tether length safety constraints. Numerical simulations were performed in Python.

\subsection{Control Barrier Function $h(x)$ Validation}
\label{h_validation}
\begin{figure}[ht]
\centering
\includegraphics[width=0.5\textwidth]{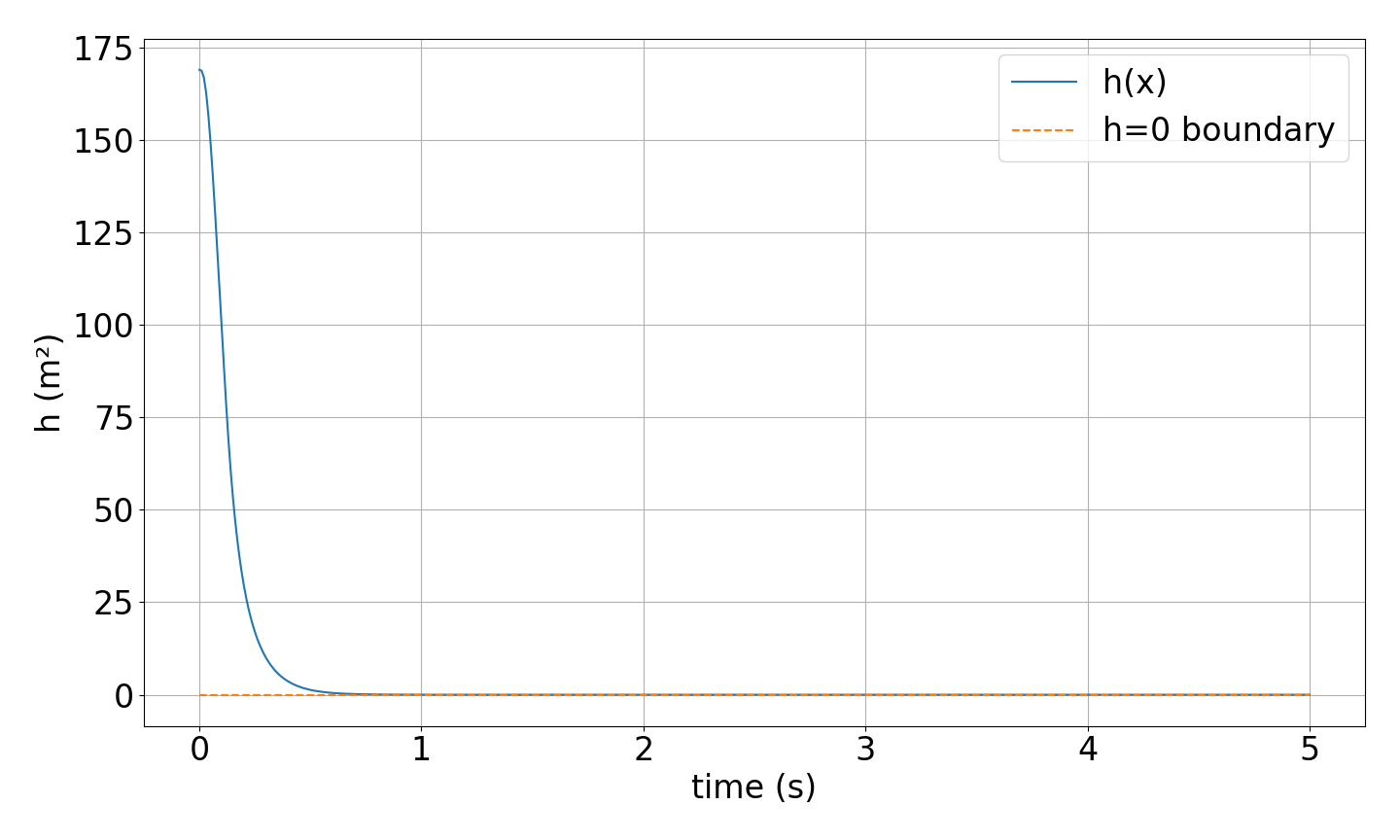}
\caption{Constraint Validation.}
\label{cbf_val}
\end{figure}

\noindent
Fig.~\ref{cbf_val} shows the evolution of the control barrier function ($h(x)$) in time . The function is designed such that maintaining $h(x) \geq 0$ ensures that the system remains within safe operational limits, in this case, preventing the UAV from exceeding the maximum length of the tether. Initially, $h(x)$ is relatively large, indicating that the UAV is well within the safe region. As the controller steers the UAV toward the target trajectory, $h(x)$ decreases but never becomes negative. By approaching but not crossing the zero boundary, the system demonstrates that the imposed safety constraint is effectively enforced.

\subsection{Tracking performances}
\begin{figure*}[ht]
\centering
\includegraphics[width=\textwidth]{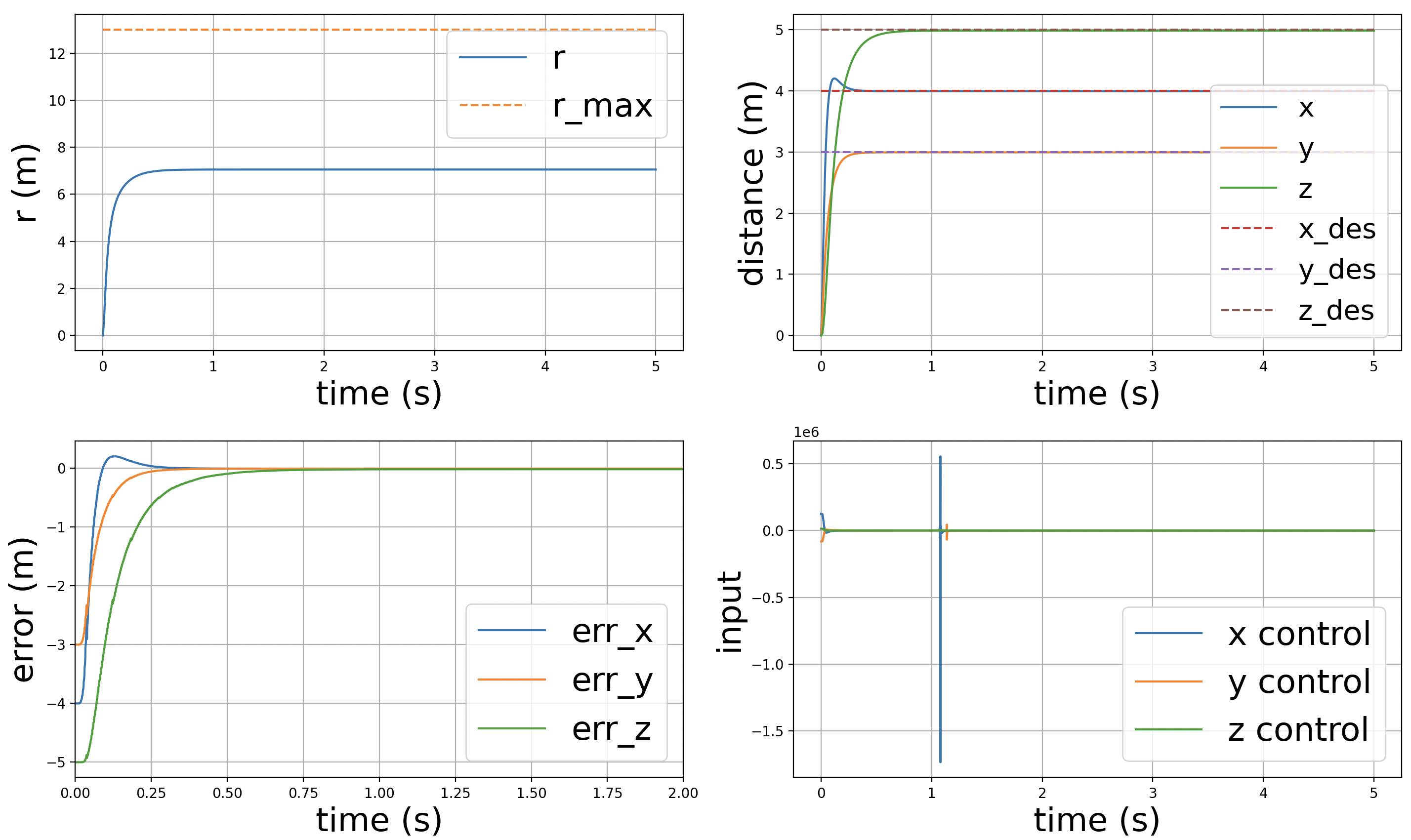}
\caption{Performance analysis of the UAV under the CBF-QP controller. The top-left plot shows the radial position \(r = \sqrt{x^2 + y^2 + z^2}\) and the maximum tether length \(r_{\text{max}}\), demonstrating compliance with the tether constraint. The top-right plot depicts the tracking of desired positions \((x_{\text{des}}, y_{\text{des}}, z_{\text{des}})\) by the UAV’s actual positions \((x, y, z)\), showing precise trajectory tracking. The bottom-left plot presents the error dynamics \(\mathbf{e}(t) = [e_x, e_y, e_z]^\top\), which converge to zero over time, indicating asymptotic stability. The bottom-right plot illustrates the bounded control inputs \((u_x, u_y, u_z)\), ensuring the feasibility of the control strategy.}
\label{fig:tracking_error_control}
\end{figure*}

Fig.~\ref{fig:tracking_error_control} shows the TUAV's performance under the proposed Control Barrier Function Quadratic Programming (CBF-QP) framework. The results confirm that the TUAV respects the tether constraint as demonstrated in Section~\ref{h_validation} and Section~\ref{constraint}, while achieving precise trajectory tracking and maintaining bounded control inputs. The radial position, defined as \(r = \|\boldsymbol{\xi}\|\), converges to a value strictly within the constraint boundary, satisfying the safety condition \(r \leq r_{\text{max}}\) at all times. The controller also achieves precise setpoint tracking, as seen in the ability of the UAV to closely follow the desired trajectory \((x_{\text{des}}, y_{\text{des}}, z_{\text{des}})\). The tracking error \(\mathbf{e}(t) = [e_x(t), e_y(t), e_z(t)]^\top = \mathbf{x}(t) - \mathbf{x}_{\text{des}}(t)\) converges to zero over time, as shown in the error dynamics. This asymptotic convergence highlights the stability of the controller and its ability to effectively minimize tracking deviations. The control inputs \((u_x, u_y, u_z)\) remain bounded throughout the simulation. This indicates that the control strategy is physically feasible, avoiding excessive control effort while ensuring compliance with safety constraints.

\subsection{Maximum tether length surface}
\label{constraint}
The results are presented to demonstrate the effectiveness of the Control Barrier Function Quadratic Programming (CBF-QP) approach in enforcing the tether constraint for a tethered unmanned aerial vehicle (UAV). The scenario considers a UAV operating around a nominal equilibrium point \((0, 0, 0)\), with a predefined maximum tether length, \(L_{\text{max}}\). The constraint on \(L_{\text{max}}\) defines a spherical region of permissible operation, geometrically represented as the surface of a sphere with radius \(L_{\text{max}}\).

\begin{figure} [ht]
\centering
\subfloat[Linear tracking scenario]{
    \includegraphics[width=0.45\textwidth]{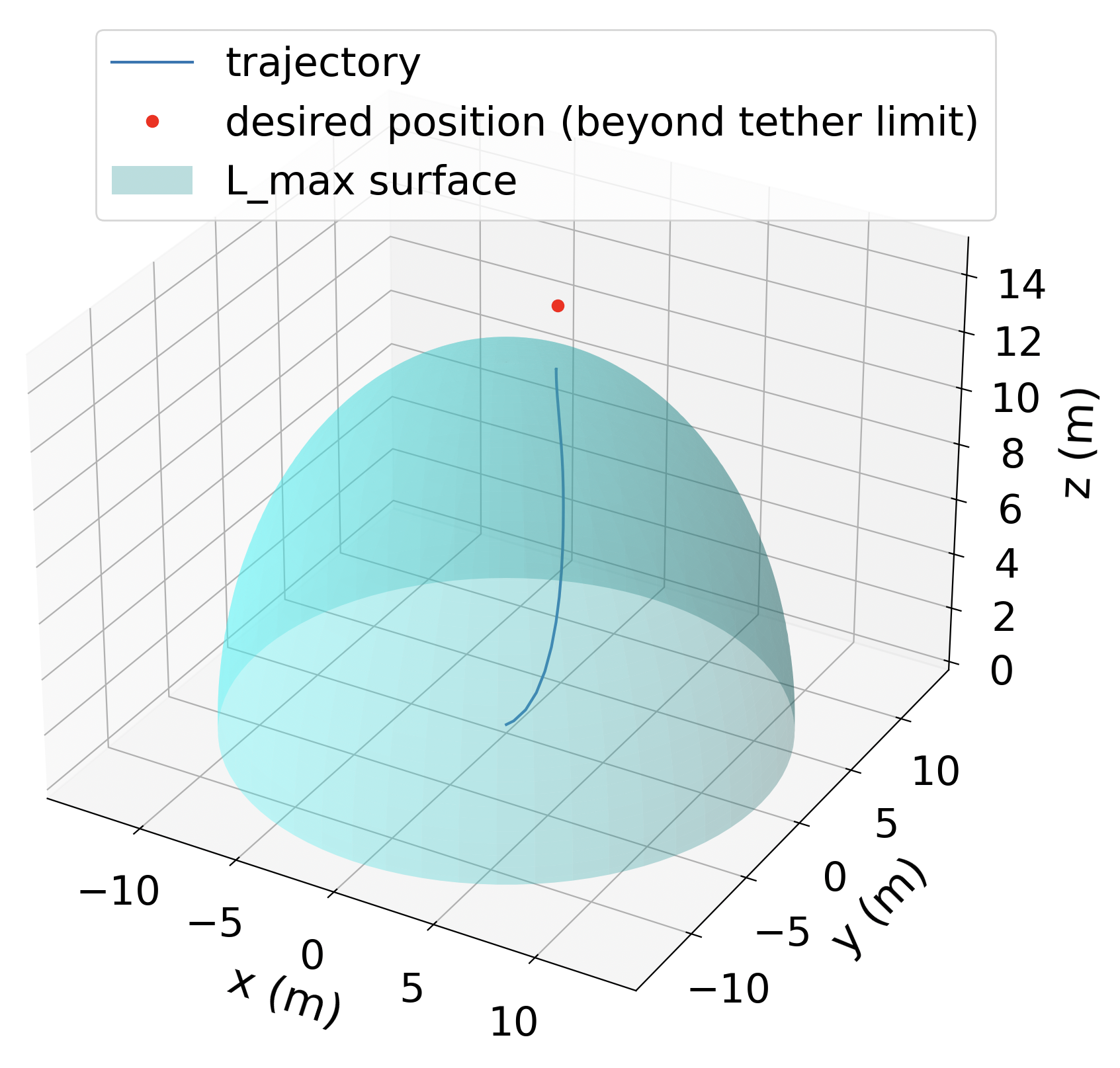}
    \label{fig:l_max_surface_linear}
}
\hfill
\subfloat[Circular tracking scenario]{
    \includegraphics[width=0.45\textwidth]{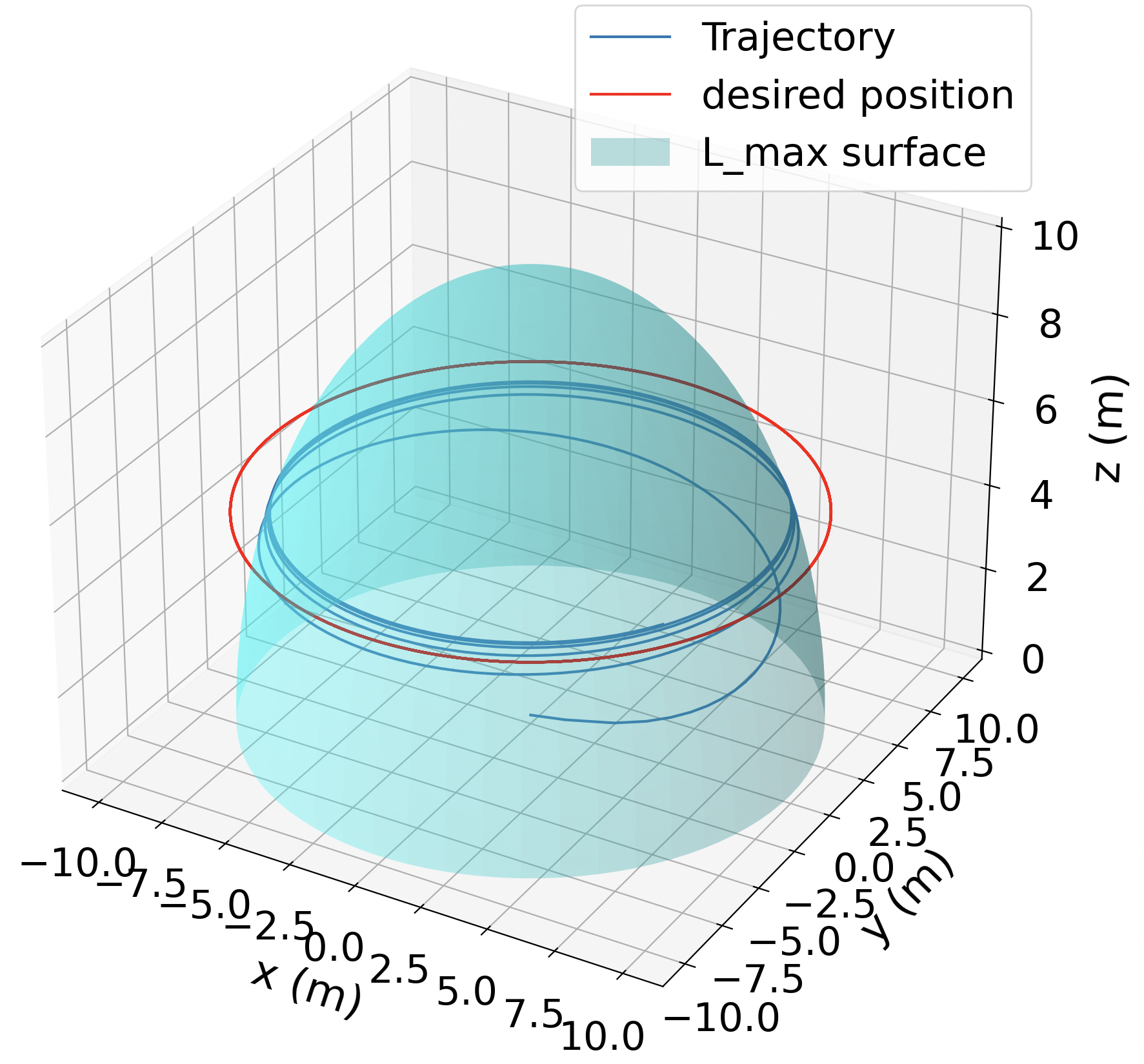}
    \label{fig:l_max_surface_circular}
}
\caption{Maximum tether length surface. The subfigures illustrate the UAV’s trajectory under (a) linear tracking and (b) circular tracking scenarios. In both cases, the UAV respects the tether length constraint defined by \(L_{\text{max}}\).}
\label{l_max_surface}
\end{figure}

To evaluate the performance of the system under the enforcement of constraints, a desired position was specified outside the permitted spherical region, beyond \(L_{\text{max}}\). As shown in Fig.~\ref{l_max_surface} (a), the trajectory of the UAV illustrates that the system dynamically respects the tether constraint. When the UAV approaches the limit defined by \(L_{\text{max}}\), the CBF-QP controller ensures that the UAV halts further progression beyond the constraint, thus enforcing the safety requirement \(||\mathbf{x}|| \leq L_{\text{max}}\). This result demonstrates the system's ability to comply with the tether constraint while minimizing deviation from the desired trajectory.

This was further verified using a circular tracking scenario, where the radius of the desired trajectory was intentionally chosen to exceed \(L_{\text{max}}\). The corresponding plot in Fig.~\ref{l_max_surface} (b) shows that the UAV respects the tether constraint while attempting to follow the circular path. These results confirm that the proposed framework ensures the adherence of the constraints across various trajectory profiles.

\section{CONCLUSION}

This paper presented a Control Barrier Function Quadratic Programming (CBF-QP) framework for ensuring safe and accurate operation of a tethered unmanned aerial vehicle (TUAV). By incorporating the tether constraint directly into the control design, the proposed approach guarantees that the UAV remains within a permissible operating region defined by the maximum tether length. The results demonstrated that the CBF-QP controller effectively enforces the safety constraint \(||\mathbf{x}(t)|| \leq L_{\text{max}}\) while achieving precise trajectory tracking.

Through a series of simulation studies, the framework was validated for a range of scenarios, including setpoint tracking and dynamic trajectory following. The system exhibited robust performance, ensuring convergence of the tracking error to zero while maintaining bounded control inputs. These results emphasize the controller's ability to achieve stability, safety, and feasibility under practical operating conditions.

\bibliographystyle{IEEEtran}

\bibliography{IEEEfull}

\section*{ACKNOWLEDGMENT}
None

\section*{REFERENCES}

\begin{IEEEbiography}[{\includegraphics[width=1in,height=1.25in,clip,keepaspectratio]{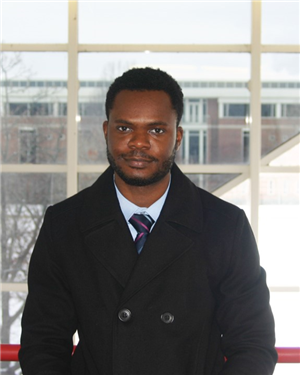}}]{SAMUEL O. FOLORUNSHO } is a Graduate member of IEEE and a PhD student conducting research at the Autonomous and
Unmanned Vehicle Systems Lab (AUVSL) and the Center for Autonomous
Construction and Manufacturing at Scale (CACMS) in the Department of
Systems Engineering at the University of Illinois, Urbana-Champaign (UIUC) under the advisorship of Prof. William R. Norris. His research is focused on control systems, computer vision and robotics - and the intersection of those for
safety-critical systems in industrial and agricultural applications.
He earned his M.S. from UIUC in 2023 and his B.S. in 2017 at
the University of Ilorin, Nigeria both in Agricultural and Biological
Systems Engineering. He has three years of working experience in management consulting.
\end{IEEEbiography}

\vspace{5pt}

\begin{IEEEbiography}[{\includegraphics[width=1in,height=1.25in,clip,keepaspectratio]{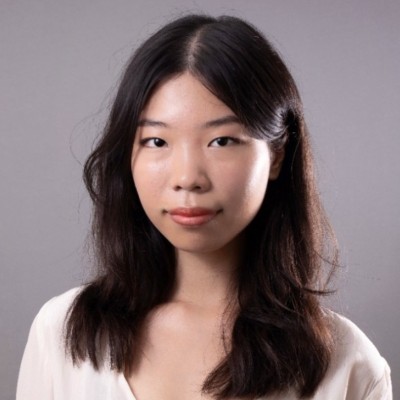}}]{MAGGIE NI } is currently pursuing the B.S. degree in Aerospace Engineering at the University of Illinois Urbana-Champaign (UIUC). She is an undergraduate research assistant at the Autonomous and Unmanned Vehicle System Laboratory (AUVSL), founded by Prof. William R. Norris. Her research interests include autonomous systems, UAV navigation, and control systems.
\end{IEEEbiography}

\vspace{5pt}

\begin{IEEEbiography}[{\includegraphics[width=1in,height=1.25in,clip,keepaspectratio]{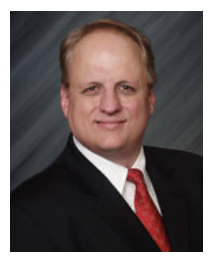}}]{WILLIAM R. NORRIS } (Member, IEEE) received
the B.S., M.S., and Ph.D. degrees in systems engineering from the University of Illinois at Urbana–
Champaign, in 1996, 1997, and 2001, respectively,
and the M.B.A. degree from the Fuqua School of
Business, Duke University, in 2007. He has over
23 years of industry experience with autonomous
systems. He is currently a Research Professor with the Industrial and Enterprise Systems
Engineering Department, University of Illinois at
Urbana–Champaign, the Director of the Autonomous and Unmanned Vehicle
System Laboratory (AUVSL), as well as the Founding Director of the Center
for Autonomous Construction and Manufacturing at Scale (CACMS).
\end{IEEEbiography}

\end{document}